# Intrinsic leakage and adsorption currents associated with the electrocaloric effect in multilayer capacitors.

M. Quintero[1,2], P. Gaztañaga[1], I. Irurzun[1]

1- Departamento de Física de la Materia Condensada, GIyA, GAIANN, Comisión Nacional de Energía Atómica, Av. Gral. Paz 1429 (1650), Buenos Aires, Argentina.
2- Escuela de Ciencia y Tecnología, Universidad Nacional de General San Martín, Martín de Irigoyen Nº 3100 (1650), Buenos Aires, Argentina.

**Abstract**

During the last few years, the increasing demand of energy for refrigeration applications has relived the interest of the scientific community in the study of alternative methods to the traditional gas-based refrigeration. Within this framework, the use of solid state refrigeration based on the electrocaloric effect reveals itself as one of the most promising technologies. In this work, we analyze how the temperature change associated with the electrocaloric effect shows a correlation with the electrical properties of a commercial multilayer capacitor. In that sense we established a clear relation between the adsorption currents and the temperature change produced by the electrocaloric effect. Additionally, intrinsic leakage currents are responsible for the sample heating due to the Joule effect. These well distinguished contributions can be useful during the design of solid state refrigeration devices based on the electrocaloric effect.

The electrocaloric effect (ECE) is the temperature change produced in a material when an external electric field produces a change in the polarization. Even when the first observation of the effect has not been categorically established [1], it is clear that the effect is well known since 1930 [2]. Some decades after that, the development of ferroelectric materials with large electric polarization gave a new boost to the topic [3,4,5,6] but the necessity of high voltages to generate an adequate electric field was an important limitation. The use of thin films allows for larger electric fields with moderated voltages [7,8,9,10,11], but the heat pump capacity of this geometry was the most significant limitation for possible solid state refrigeration devices.

The use of multilayer capacitors (MLC) was proposed as one possible solution for the above mentioned issues [12]. MLCs are interdigitated metallic electrodes containing some dielectric material in between. Thanks to this particular geometry, it is possible to reach high electric field values with moderated voltages, keeping, at the same time, a good thermal link between the active material and the external medium.

A commercial MLC based on Y5V dielectric with Ni electrodes was used to demonstrate how real the possibility of applying these systems could be, even when they were not specifically designed for this kind of applications [13]. The presence of Joule heating due to leakage currents was analyzed and decoupled from the ECE contribution to the temperature change [14]. It was established that even for small currents, the Joule heating contribution must be considered during the evaluation of cooling device efficiency based on the ECE in MLCs.

After the application of a voltage step on an MLC, an electric current flow is observed between the electrodes. Two contributions can be distinguished; one, associated with charge adsorption ($I_{AD}$), and the other, with the charge injection from metal electrodes ($I_{il}$).

The charge adsorption includes a variety of physical mechanisms such as dielectric relaxation processes, dipole orientation, redistribution of ionic charges, charge injection from the electrodes and electron tunneling into traps in the dielectric [15,16]. These currents decrease relatively slowly over time. On the other hand, intrinsic leakage currents are associated with the conduction mechanism and are not as function of time [17,18,19,20,21,22,23].

In this work, we will present direct measurements of ECE in a commercial MLC, as a function of the applied voltage.

Finally, a detailed electric characterization will be outlined to separate adsorption and leakage currents. Characteristic ECE parameters will be mapped to each of these two contribution sources.

We studied a commercially manufactured MLC[24], formed by interdigitated nickel electrodes in a doped BaTiO3 dielectric (Y5V). Even when the manufacturer recommends a maximum applied voltage of 10 V we did not find evidence of dielectric breakdown in our working range (210 V). In all the measurements the values of leakage currents remain below 1uA, far from the expected in a dielectric breakdown[25].

The MLC was attached to a platinum thermometer (Pt-1000) using vacuum grease to ensure good thermal link between both elements. The system was placed on a Cu block using a 2mm Teflon layer as thermal insulation. The block was placed in a vacuum chamber, with a pressure of 10-5 Torr. The electrical characterization of the sample was performed using a Keithley 2400 source and the resistance of the Pt-1000 thermometer was measured using an Agilent 3401 multimeter. All measurements were performed at room temperature.

In figure 1b we show the excess temperature registered by the thermometer during the application of 210 V at room temperature. After the voltage application, the temperature rapidly increases due to the ECE, followed by a slow relaxation. A sudden decrease in temperature is observed when the voltage is turned off, followed by a slow relaxation and return to the initial temperature.

The temperature evolution is a consequence of the heat exchange between the different parts of the experimental setup and can be depicted through the corresponding Tian´s equations [26]

$$\frac{dQ_S}{dt} = \gamma_{SH}(T - T_H) + \gamma_{ES}(T - T_0) = -C_S \frac{dT}{dt} + iV + \zeta(V)\frac{dV}{dt}$$

$$\frac{dQ_H}{dt} = \gamma_{SH}(T_H - T) + \gamma_{EH}(T_H - T_0) = -C_H \frac{dT_H}{dt} + P_{th}$$

The first equation describes the heat exchange balance in the sample (including the ECE contribution), and the second takes into account the heat exchange balance in the sample holder.

T is the temperature of the MLC and $T_H$ is the temperature of the sample holder, being the last one the temperature measured by the thermometer. $T_0$ is the temperature of the thermal bath and V is the voltage applied between the electrodes of the MLC.

$\gamma_{ES}$ is the thermal conductance between the sample and the thermal bath. $\gamma_{EH}$ is the thermal conductance between the thermometer and the thermal bath.

$\gamma_{SH}$ is the thermal conductance between the MLC and the sample holder.

The temperature change due to the ECE is represented by the function $\zeta(V)$ and $P_{th}$ is the Joule effect dissipation due to the thermometer. $C_S$ and $C_H$ are the heat capacities of the sample and the holder, respectively.

The temperature response of the system after the voltage is applied can be modeled by

$$T_H = T_0 + T^* + A_1 e^{-\frac{t}{\tau_{SH}}} + A_2 e^{-\frac{t}{\tau_E}}$$

where T* is the temperature of stabilization, and $\tau_{SH}$ and $\tau_E$ are the characteristic times for the heat exchange between the MLC and the Pt-1000 and between the MLC and the Pt-1000, and the thermal bath.

It is possible to establish a relation between the parameters $A_1$ and the adiabatic temperature change of the sample due to the ECE. T* is the temperature reached in the steady state, when there is no contribution from the ECE. The value of T* is then related with the Joule effect power dissipation and with the thermal link between the sample and the thermal bath [14] ($T^* \approx \frac{iV}{\gamma_{ES}}$).

In order to understand how the electrical behavior of the MLC is related with the temperature change, we performed a careful characterization of the electrical properties. After changing the voltage between the electrodes an electrical current appears due to the change produced in the boundary conditions of the dielectric and due to the charge injection from the electrodes. Since both contributions present a strong dependence on the voltage, the first one decreases over time while the second one remains constant. We used this distinction to separate both contributions to the total current. The voltage was increased in small steps (5V), and the current was measured during a long time (200 sec.) keeping V constant. In figure 2, we present the time dependence on the current during the abovementioned experiment.

For each value of V, we obtained the time evolution of the total current (inset of figure 2).

In all cases, we observed that the current tends to a stable value corresponding to the intrinsic leakage current ($I_{il}$). Subtracting $I_{il}$ from the signal, it is possible to obtain the time evolution of adsorption current ($I_{AD}$). Integrating $I_{AD}$ in time, we obtain the charge adsorbed during relaxation.

In figure 3, we show the variation of $I_{il}$ depending on the applied voltage. The current remains close to zero below 100 V. Above 100 V, an $I_{il}$ increase following increases in voltage is observed, and the dependence is consistent with electrode limited Schotky emission or bulk limited Poole-Frenkel transport, in line with the reported findings in other MLCs[27].

In figure 4, we show the adsorbed charge ($Q_{AD}$) as a function of the applied voltage during the abovementioned experiments. Starting from an uncharged state, we observed an increase in the charge reaching a value of 4μC at the maximum applied voltage (200 V). Upon decreasing the voltage, a remnant charge of 1μC remains in the MLC, and a voltage of around 20 V in the opposite sense is necessary to discharge the capacitor. If we continued with the voltage variation, a hysteretic behavior is consistent with the polarization loops observed in similar systems[28].

The above presented analysis enables us to separate two contributions to the current between the electrodes on the MLC. We will focus on how each of these contributions has an impact on the temperature variation of the system.

In figure 5a, we present the evolution of the T* parameter as a function of the power dissipated by the intrinsic leakage current. The linear dependence observed is indicative that this current is the one responsible for the Joule heating observed in the sample for large voltages. Taking into account the dependence of $I_{il}$ on the voltage, we can consider a low dissipation region below 100 V, this condition being the most suitable for further use in applications.

On the other hand, in figure 5b, we present the value of the $A_1$ parameter as a function of the absorption charge. The linear relation observed is indicative that the absorption and the associated charge are strongly related to the ECE in the sample. As the entropy decreases with polarization, a positive correlation between the adsorption charge and the magnitude of ECE is expected.

Summarizing, we studied the electrical properties of a commercial multilayer capacitor in order to understand how these properties are related to the temperature change associated with the electrocaloric effect. When voltage is applied between the electrodes of the MLC, two different contributions to the current are observed: one associated with the intrinsic leakage ($I_{il}$) and the other one related to the adsorption ($I_{AD}$). In the case of $I_{il}$, we observed voltage dependence consistent with mechanisms of conduction previously observed in similar samples. By means of the $I_{AD}$ integration, we observed a hysteretic behavior of the $Q_{AD}$ as a function of the applied voltage. This kind of dependences reminds us of the polarization observed in other samples with similar characteristics. Comparing the values of the abovementioned parameters with those extracted from the direct measurement of the sample temperature during the experiment, we could establish a relation between $I_{il}$ and the Joule heating, as well as between $Q_{AD}$ and the electrocaloric effect. This information could be extremely useful during the design of a cooling device based on the ECE, both to avoid Joule heating and to develop an optimized geometry aimed at maximizing the "usable" temperature change.

This work has been done with the support of ANPCyT PICT 1506/2012. M.Q. is also member of CIC-CONICET.

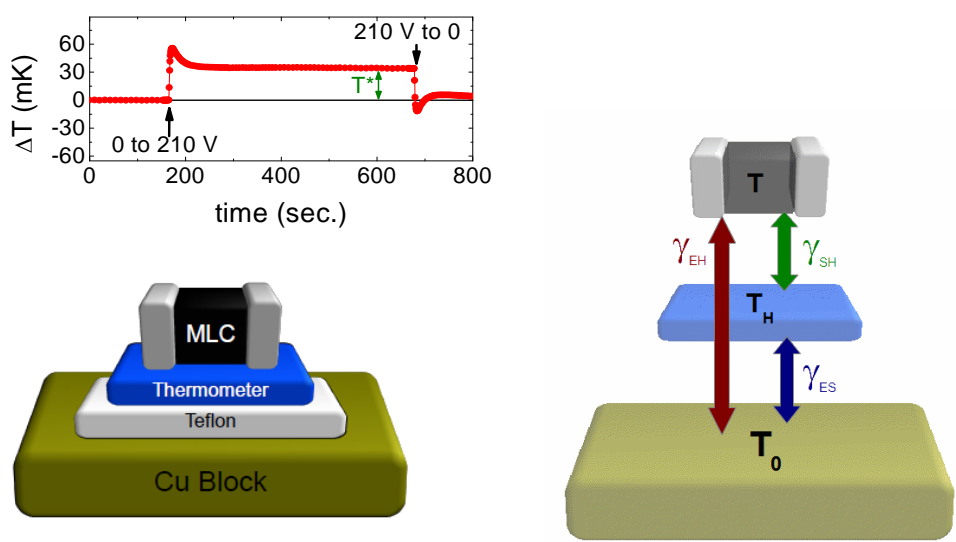

Figure 1: (Top) Typical curve of sample temperature as a function of time during the application of 210V. (Bottom) Schematic representation of the experimental setup for the direct measurement of the temperature change.

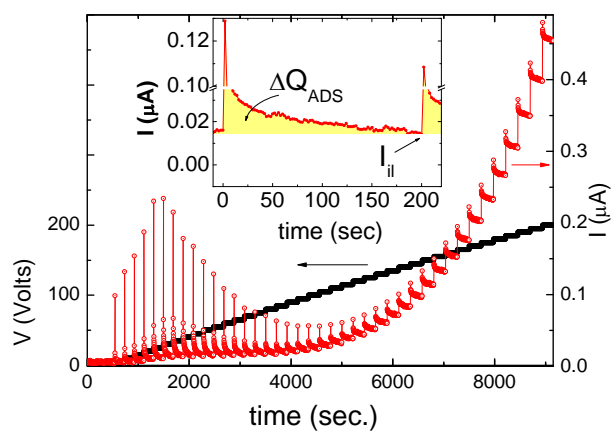

Figure 2: Time dependence of the measured current during the application of increasing voltage in steps of 5V. The inset shows the enlargement of the current measured in one step.

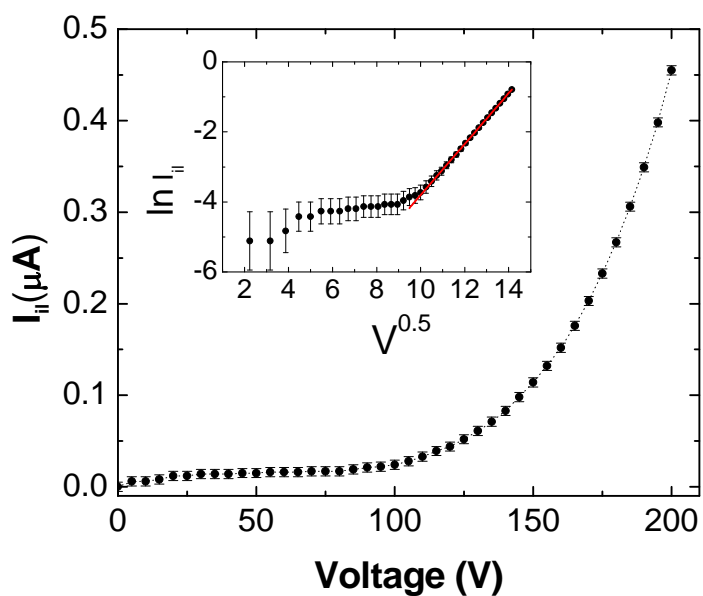

Figure 3: Intrinsic leakage current (Iil) as a function of the applied voltage. The inset shows the same data as a function of $V^{0.5}$. The linear behavior above 100 V is consistent with a Poole-Frenkel dependence.

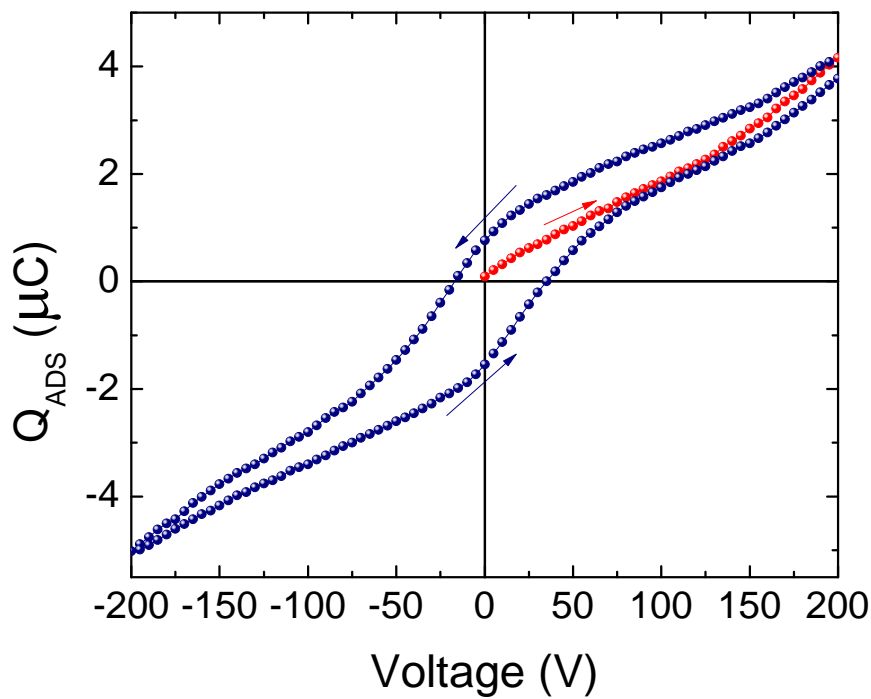

Figure 4: Adsorption charge as a function of the applied voltage (steps of 5V).

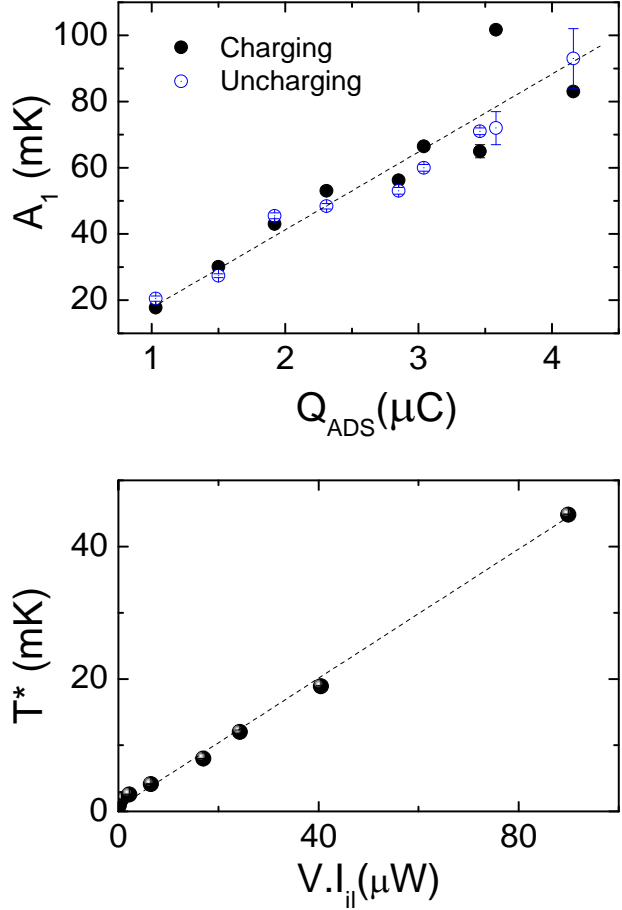

Figure 5: (Top) Parameter A1 as a function of the adsorption charge. (Bottom) Parameter T* as a function of the power dissipated by Joule heating produced by the intrinsic leakage current.